\documentclass[aps,preprint,showpacs,showkeys,superscriptaddress]{revtex4-1}

\usepackage{dcolumn}% Align table columns on decimal point
\usepackage{bm}% bold math
\usepackage{amssymb}
\usepackage{amsfonts}
\usepackage{amsmath}
\usepackage{graphicx}
\usepackage{epstopdf}
\usepackage{float}
\usepackage{relsize}
\usepackage{mathtools}
\usepackage{placeins}
\usepackage{array}
\usepackage{xcolor}
\begin{document}

\title{Regular black holes from Kiselev anisotropic fluid}% Force line breaks with 

\author{L. C. N. Santos}
\email{luis.santos@ufsc.br}

\affiliation{Departamento de F\'isica, CFM - Universidade Federal de Santa Catarina; \\ C.P. 476, CEP 88.040-900, Florian\'opolis, SC, Brazil.}

\begin{abstract}
In this paper, we investigate a generalization of Kiselev black holes by introducing a varying equation of state parameter for the anisotropic fluid surrounding the black hole. We extend this model by allowing $w$ in the expression $p_t(r)/\rho(r) = (3w + 1)/2$ to vary as a function of the radial coordinate, and derive new solutions to the Einstein field equations for this configuration. In particular, we study solutions that describe regular black holes. By choosing specific forms of  $w(r)$, we obtain regular black hole solutions, and show that the matter surrounding the black hole can satisfy the weak and strong energy conditions under certain values of parameters analyzed. Due to the generality of this treatment, other categories of black holes can be obtained with particular choices of the parameter of equation of state. Our analysis confirms that the curvature invariants associated with the regular black holes remain finite at the origin, indicating the absence of singularities. We also explore the physical properties of the matter associated with these solutions. Due to the versatility, we suggest the possibility of using this approach as a tool to construct new physical solutions associated with regular black holes or other geometries of interest. 
\end{abstract}


%\pacs{04.20.−q }
\maketitle

\preprint{}

\volumeyear{} \volumenumber{} \issuenumber{} \eid{identifier} \startpage{1} %
\endpage{}
\section{Introduction}
The concept of black holes, arising from the predictions of general relativity and modified theory of gravity of, has evolved into a fundamental research topic of physics and astrophysics. The most common of these compact objects, characterized by their event horizons, have a point singularity at the center, where spacetime curvature becomes infinite. This singularity represents a breakdown in the structure of spacetime. Some studies suggest that singularities at the cores of black holes demand a quantum theory of gravity for a complete resolution \cite{hawking1975particle, penrose1965gravitational}. Despite the open questions about the nature of the solutions of Einstein field equations containing singularities, we have recently seen many predictions of the theory being confirmed. The detection of gravitational waves \cite{waves1} by the Virgo and LIGO (Laser Interferometer Gravitational-Wave Observatory) research teams stands as one of the most significant discoveries in physics this century. Since then, the Virgo, LIGO, and KAGRA (Kamioka Gravitational Wave Detector) international collaborations have collectively detected over 200 gravitational wave events \cite{abbott2023gwtc}. Initially predicted by Albert Einstein in 1916, gravitational waves are a direct consequence of the field equations in General Relativity. In 2019, another major prediction of General Relativity was confirmed: the first direct image of a black hole, particularly the accretion disk and the area around the event horizon \cite{black1}.

It is possible to address the problem of singularities within the classical physics. Bardeen proposed a black hole model \cite{bardeen1968non} that avoids the singularity by replacing the mass
of Schwarzschild black hole with a function that depends on radial coordinate.  Subsequent works considered regular black holes constructed using various theoretical approaches, including coupling Einstein equations with nonlinear electrodynamics \cite{ayon1998bardeen,bronnikov2001regular,han2020thermodynamics,bronnikov2020comment,junior2024regular,walia2024exploring,dolan2024superradiant,tangphati2024magnetically,guo2024recovery,kar2024novel,bronnikov2024regular}. Models for regular rotating black holes can be constructed, for instance, considering the Hayward and to the Bardeen black hole metrics. In this context, the Newman–Janis algorithm was employed in obtaining a rotating black hole \cite{bambi2013rotating,toshmatov2017generic}. However, it was later shown that the standard Newman–Janis algorithm can lead to inconsistencies in generating rotating solutions coupled to nonlinear electrodynamics \cite{kubizvnak2022slowly}. Additionally, recent observations have aimed at testing candidates for regular black holes. For example, using the shadows of the black hole M87*, it was not possible to determine whether rotating black holes are preferential when compared with observational data \cite{kumar2020testing}. 
Regular black holes can be constructed in two distinct ways. The first way consists of solving the field equations of classical gravity by considering an energy-momentum tensor associated with a particular distribution of matter. The other way is to derive regular black holes as quantum corrections to classical singular black holes. Such corrections can come, for example, from the effects of loop quantum gravity \cite{modesto2004disappearance,gambini2008black,ashtekar2023regular,modesto2006loop,momennia2022quasinormal,sharif2010quantum}.

Considering black holes with an energy-momentum tensor associated with a particular configuration of matter, a type of black hole surrounded by an fluid that has received recent attention is the Kiselev black hole \cite{kiselev}, also called  black holes with quintessence in some works in the literature. Several physical systems have been studied using this particular equation of state for the anisotropic fluid \cite{santos2023kiselev}.  Particular values of the parameter of the equation of state $w$  can reproduce, the cosmological accelerating pattern observed \cite{kiselev}. Besides, Kiselev anisotropic fluid have been studied in the context of shadow of black holes \cite{shadow1,shadow2,shadow3,shadow4}, quasinormal modes \cite{quasi1,quasi2,quasi3,quasi4,quasi5,quasi6,quasi7,quasi8,quasi9}, thermodynamics of black holes \cite{termo1,termo2,termo3,termo4,termo5,termo6,termo7,termo8}, and in modified gravity \cite{heydarzade2017black,sakti_kerrnewmannutkiselev_2020,morais_thermodynamics_2022,santos2023kiselev,gogoi2023joule,ghosh_rotating_2024}. Additional properties of the Kiselev black hole and  the collapse of null and timelike thin shells within this spacetime was considered in \cite{saadati2021thin,javed2024impact,javed2024klein}. 

In particular, Kiselev equation of state connects energy density and radial pressure in the form $p_t(r)/\rho(r) = (3w + 1)/2$, where $w$ is a constant. In this work, we generalize this equation of state considering a varying equation of state parameter $w(r)$. We consider the Einstein field equation with a stress-energy tensor associated with anisotropic fluid. Then we obtain a general solution that depends on the Kiselev parameter $w(r)$. In our work, we show that several news solutions of field  equations can be obtained by specifying particular form of the function $w(r)$. We test our approach considering regular black holes. By choosing particular forms of the function $w(r)$, we obtain a solution of the field equation associated with regular black holes. The calculation of the curvature-invariant quantities indicates that such geometries are indeed free of singularities at the origin. 

The paper is organized as follows: In Section II, we consider a spherically symmetric spacetime with the line element corresponding to a static geometry and solve the Einstein field equation in the presence of an anisotropic stress-energy tensor obtaining a general solution. In Section III, we discuss particular solutions associated with regular black holes in general relativity. By choosing particular shapes for $w(r)$, we obtain examples of geometries without singularity. we investigate the characteristics of the matter associated with the obtained regular solutions through the energy conditions in Section IV. Finally, in Section V, are our concluding comments. In this paper, we use natural units where $c = G= 1$.

\section{Einstein field equations: the generalized solution}
Let us consider Einstein field equations in general relativity. The first step in solving these equations is to consider initial geometric properties of spacetime that include certain symmetries. A spherically symmetric spacetime with the line element corresponding to a static geometry with two general radial functions depending on position coordinates can be written as  
\begin{equation}
    ds^2 = B(r) dt^2 - A(r) dr^2 - r^2(d\theta^2 + \sin^2\theta \, d\phi^2),
    \label{21}
\end{equation}
where \( B(r) \) and \( A(r) \) are functions of the radial coordinate \( r \).  
The energy-momentum tensor for Kiselev black holes is defined such that the spatial components are proportional to the time component:
\begin{align}
    T^{t}_{\:\:\:t} = T^{r}_{\:\:\:r} &= \rho(r), \label{22} \\
    T^{\theta}_{\:\:\:\theta} = T^{\phi}_{\:\:\:\phi} &= -\frac{1}{2} \rho(r) (3w + 1),
    \label{23}
\end{align}
where \( w \) is the equation of state parameter.  
By averaging over the angles isotropically, in place of equations (\ref{22}) and (\ref{23}), one arrives at the barotropic equation of state. Here, we propose to consider that $w$ is a variable that depends on $r$, i.e., $w=w(r)$, generalizing the original Kiselev equation of state. For simplicity, let us define the following function 
\begin{equation}    
\Bar{w}(r) \equiv \frac{1}{2}(3w(r)+1). 
\end{equation}
From now on, we will use this expression in our calculations.
Note that in Kiselev black holes \cite{kiselev}, the components of the energy-momentum tensor are effectively related to an anisotropic fluid, given by
\begin{equation}
    T^{\mu}_{\:\:\:\nu} = \text{diag}(\rho, -p_r, -p_t, -p_t),
    \label{24}
\end{equation}
with \( p_r = -\rho \) and \( p_t = \frac{1}{2} \rho (3w + 1) \), which can be derived from the general form of an anisotropic fluid \cite{santos00}:
\begin{equation}
    T_{\mu\nu} = -p_t g_{\mu\nu} + (p_t + \rho) U_{\mu} U_{\nu} + (p_r - p_t) N_{\mu} N_{\nu},
    \label{25}
\end{equation}
where \( p_t(r) \), \( \rho(r) \), and \( p_r(r) \) represent, respectively, the tangential (or transverse) pressure, energy density, and radial pressure of the fluid. The quantities \( U_{\mu} \) and \( N_{\mu} \), which denote the four-velocity and the radial unit vector, are defined as
\begin{equation}
    U^{\mu} = \left( \frac{1}{\sqrt{B(r)}}, 0, 0, 0 \right),
    \label{26}
\end{equation}
\begin{equation}
    N^{\mu} = \left( 0, \frac{1}{\sqrt{A(r)}}, 0, 0 \right),
    \label{27}
\end{equation}
satisfying the conditions \( U_{\nu} U^{\nu} = 1 \), \( N_{\nu} N^{\nu} = -1 \), and \( U_{\nu} N^{\nu} = 0 \). 
Considering the Einstein field equations
\begin{equation}
    G^{\mu}_{\:\:\:\nu}=R^{\mu}_{\:\:\:\nu}-\frac{1}{2}\delta^{\mu}_{\:\:\:\nu}R = 8\pi T^{\mu}_{\:\:\:\nu},
    \label{28}
\end{equation}
the spherically symmetric spacetime  corresponding the line element (\ref{21}) and the energy-momentum tensor related to an anisotropic fluid (\ref{24}) imply that Einstein field equations (\ref{28}) can written as 
\begin{align}
    G^{t}_{\:\:\:t}&=8\pi\rho(r) \label{29},\\
    G^{r}_{\:\:\:r}&=8\pi \rho(r) \label{210}, \\
    G^{\theta}_{\:\:\:\theta}&=-8\pi \Bar{w}(r)\rho(r) \label{211},
\end{align}
where

\begin{align}
  G^{t}_{\:\:\:t}&=\frac{\frac{d}{dr}A(r)}{rA(r)^2} + \frac{1}{r^2} - \frac{1}{r^2 A(r)}
 \label{212},\\
    G^{r}_{\:\:\:r}&= - \frac{\frac{d}{dr} B(r)}{r A(r) B(r)} + \frac{1}{r^2}  - \frac{1}{r^2 A(r)}
\label{213}, \\
     (F(r))G^{\theta}_{\:\:\:\theta}&= -2 \left( \frac{d^2}{dr^2} B(r) \right) A(r) B(r) r \nonumber\\ 
&+ \left( \frac{d}{dr} B(r) \right)^2 A(r) r \nonumber\\
&+ B(r) \left( \left( \frac{d}{dr} A(r) \right) r - 2 A(r) \right) 
\left( \frac{d}{dr} B(r) \right) \nonumber\\
&+ 2 B(r)^2 \left( \frac{d}{dr} A(r) \right)
\label{214},   
\end{align}
with $F(r) = 4A(r)B(r)^2 r$. 
By taking into account the original Kiselev approach \cite{kiselev}, Eqs. (\ref{212}) and (\ref{213}) yield the relation 
\begin{equation}
    G^{t}_{\:\:t}=G^{r}_{\:\:r}
     \label{215}.
\end{equation}
 Then, the symmetry arising from this expression demands  
\begin{equation}
    B(r)\frac{dA(r)}{dr}+A(r)\frac{dB(r)}{dr}=0
    \label{216},
\end{equation}
with solution in the form $A(r)=1/B(r)$. This imply that the original set of field equations can be written as
\begin{align}
   G^{t}_{\:\:t}=G^{r}_{\:\:r}=& -\frac{1}{r}\frac{dB(r)}{dr} - \frac{B(r)}{r^2} + \frac{1}{r^2}= 8\pi\rho, \label{217}\\
 G^{\theta}_{\:\:\theta}=G^{\phi}_{\:\:\phi}=& -\frac{1}{2}\frac{d^2B(r)}{dr^2}-\frac{1}{r}\frac{dB(r)}{dr} = -8\pi\Bar{w}(r)\rho
    \label{218}.
\end{align}
At this point, we have reduced the original set of differential equations to Equations (\ref{217}) and (\ref{218}). Now, energy density can be eliminated and Equations (\ref{217}) and (\ref{218}) are combined in the form
\begin{equation}
    - \frac{d^2 B}{dr^2} r^2 - 2r \left( \Bar{w}(r) + 1 \right) \frac{dB}{dr} - 2 \Bar{w}(r) (B - 1) =0,
    \label{219}
\end{equation}
despite the presence of function $\bar{w}(r)$, Eq.  (\ref{219}) can be integrated, and the result is written as 
\begin{equation}
    B(r) = \frac{C_1}{r} \int \frac{dr}{\exp\left(\int \frac{\Bar{w}(r)}{r} \, dr\right)^2}  + 1 + \frac{C_2}{r},
    \label{220}
\end{equation}
where $C_1$ and $C_2$ are integration constants. If $C_2 = -2M$, where $M$ is the ADM mass of the black hole in the absence of fluid, then this term is associated with usual Schwarzschild black hole, in contrast, the integral encompasses the effect of fluid. This expression is formally a general solution for an anisotropic fluid with varying equation of state parameter $\Bar{w}(r)$.  The integral term can be solved by choosing an expression for $w(r)$ that depends on the particular form of the solution sought. In the particular case in which $w$ is a constant, we obtain the solution
\begin{equation}
  B(r) = 1 + \frac{C_2}{r} + \left(\frac{D}{r}\right)^{3w+1},  
\end{equation}
that is the original Kiselv metric function \cite{kiselev}. In the next sections, we can analyze Eq. (\ref{220}) by considering solutions associated with regular black holes.

\section{Regular black holes from varying equation of state parameter}
Let us explore particular forms of the function $\Bar{w}(r)$ associated with regular black holes in general relativity. The first case that will be chosen corresponds to the function
\begin{equation}
    \Bar{w}(r)  = -\frac{2a^2-3r^2}{2(a^2 + r^2)}.
    \label{31}
\end{equation}
By substituting this (\ref{31}) in (\ref{220}), we obtain the result
\begin{equation}
    B(r) =1 +\frac{C_2}{r}- \frac{C_3r^2}{(a^2 + r^2)^{3/2}},
    \label{32}
\end{equation}
where $C_3 \equiv -C_1/3a^2$. By choosing $C_2 = 0$, Equation (\ref{32}) is associated with a regular space-time. This can be seen from the curvature invariants 
\begin{align}
  R &=  -3\frac{(4a^2 - r^2) C_3 a^2}{(a^2 + r^2)^{7/2}},\label{33} \\
  R_{\alpha\beta}R^{\alpha\beta} &=\frac{9 \left( 8a^4 - 4a^2r^2 + 13r^4 \right)^2 C_3^2 a^4}{2 \left( a^2 + r^2 \right)^7}, \label{34} \\ R_{\alpha\beta\gamma\delta}R^{\alpha\beta\gamma\delta} & = \frac{3C_3^2 \left( 8a^8 - 4a^6r^2 + 47a^4r^4 - 12a^2r^6 + 4r^8 \right)}{(a^2 + r^2)^7}, 
  \label{35}
\end{align}
which are regular everywhere. In the case of Kiselev anisotropic fluid, the function in Equation (\ref{31}) can be related to the original Kiselev varying equation of state parameter $w(r)$ as 
\begin{align}
    w(r) &= \frac{2}{3}\Bar{w}(r) - \frac{1}{3} \nonumber\\
     & = -\frac{2(2a^2-3r^2)}{6(a^2 + r^2)}- \frac{1}{3}
     \label{36}.
\end{align}
%%%%%%%%%%%%%%%%%%%%%%%%%%%%%%%%%%%%%%%%%%
\begin{figure}[H]
\centering
\includegraphics[scale=0.55]{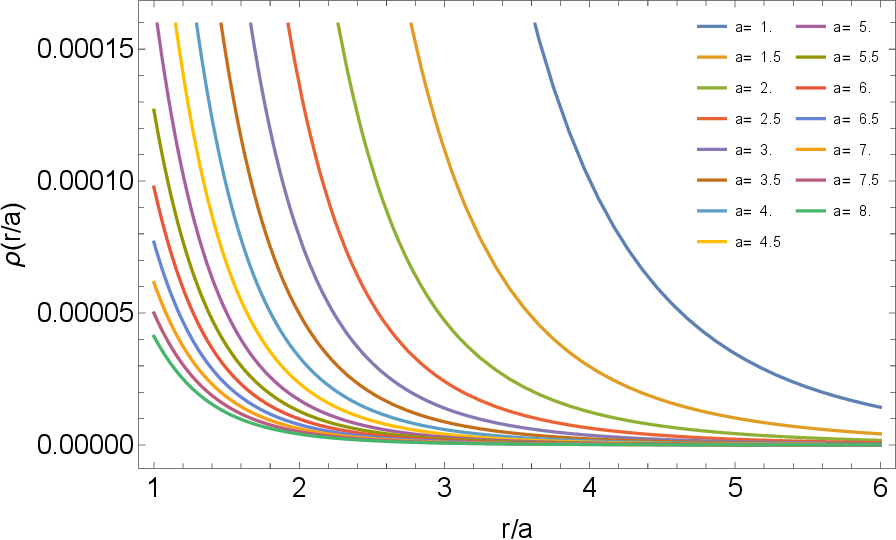} 
\caption{Plot of the energy density as a function of $r$ for the regular solution associated with $w(r)$ given in Equation (\ref{31}). We choose several values for $a$ using $C_3 = 1$.   
}
\label{f3}
\end{figure}
%%%%%%%%%%%%%%%%%%%%%%%%%%%%%%%%%%%%%%%%%%
We can see that by considering the asymptotic limit, $w(r)$ goes as 2/3. On the other hand, substituting Equation (\ref{32}) in the field equation, we obtain the following expression for the energy density  
\begin{equation}
   \rho(r) =  \frac{3 C_3 a^2}{8\pi(a^2 + r^2)^{\frac{5}{2}}},
    \label{37}
\end{equation}
that is the energy density associated with the surrounded Kiselev anisotropic fluid. In Figure \ref{f3}, we construct a graph with various values for $a$ illustrating the behavior of the energy density as a function of the coordinate $r$. As can be seen, the energy density tends to values close to zero for large $r$. Solution (\ref{32})  is similar to the Bardeen model obtained in the context of nonlinear electrodynamics. Here, we interpret solution (\ref{32}) as effect of a particular configuration of the anistropic fluid. 

Another class of solutions associated with regular black holes can be obtained by considering the following choice for the function of the equation of state
\begin{equation}
    \Bar{w}(r) = -\frac{2C_4 - 3r^3}{2C_4}
     \label{38},
\end{equation}
using this expression in Equation (\ref{220}), we obtain the following black hole function
\begin{equation}
    B(r) = 1 + \frac{C_2}{r} -\frac{C_1 C_4/3}{r \exp\left(\frac{r^3}{C_4}\right)},
    \label{39}
\end{equation}
where $C_1,C_2$ and $C_4$ are constants. The energy density for this solution is given by
%%%%%%%%%%%%%%%%%%%%%%%%%%%%%%%%%%%%%%%%%%
\begin{figure}[H]
\centering
\includegraphics[scale=0.58]{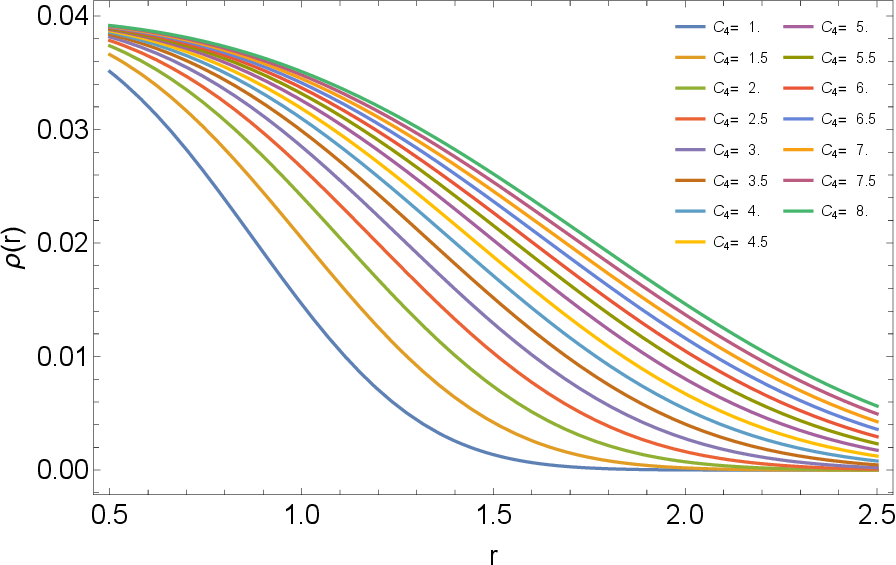} 
\caption{Plot of the energy density as a function of $r/a$ for the regular solution associated with $w(r)$ given in Equation (\ref{38}). We choose several values for $C_4$ using $C_1 = -1$   
}
\label{f4}
\end{figure}
%%%%%%%%%%%%%%%%%%%%%%%%%%%%%%%%%%%%%%%%%%
\begin{equation}
    \rho(r) = -\frac{C_1\exp{\left(-\frac{r^3}{C_4}\right)}}{8\pi}.
    \label{300}
\end{equation}
In this way, due to the presence of the exponential function term, the energy density has different behavior near the origin. In Figure \ref{f4}, we construct a graph with various values for $a$ illustrating the behavior of the energy density as a function of the coordinate $r$. As can be seen, the energy density tends to values approaching zero for large $r$. By redefining constants, function (\ref{39}) can be written as 
\begin{equation}
    B(r) = 1 - \frac{r_g}{r}\left(1-\exp\left(-\frac{r^3}{r_0^3}\right) \right), 
    \label{301}
\end{equation}
 This solution \cite{dymnikova1992vacuum} has interesting properties. For large $r$, the exponential term is negligible and the solution is similar to the Schwarzschild solution. Furthermore, unlike the solution given by equation (\ref{32}), it is regular even with the presence of the Schwarzschild term $r_g/r$. The curvature invariants for the metric function (\ref{301}) give us the expressions
\begin{align}
    R & =  \exp\left(-\frac{r^3}{r_0^3}\right) \left(3 r^3 - 4 r_0^3\right) \frac{3 r_g}{r_0^6},
 \label{302} \\   R_{\alpha\beta}R^{\alpha\beta} & = \frac{9 \exp\left(-\frac{r^3}{r_0^3}\right)^2 r_g^2 \left(9 r^6 - 12 r^3 r_0^3 + 8 r_0^6\right)}{2 r_0^{12}},
 \label{303}\\ R_{\alpha\beta\gamma\delta}R^{\alpha\beta\gamma\delta} & = (12 r_g^2) \frac{1}{r^6} + \nonumber\\
 &\Bigg( 
-(36 r_g^2) \frac{1}{r_0^6} - (24 r_g^2) \frac{1}{r_0^3} \frac{1}{r^3} \nonumber\\
& - (24 r_g^2) \frac{1}{r^6} \Bigg) \frac{1}{\exp\left(\frac{r^3}{r_0^3}\right)} \nonumber\\
& + \Bigg( 
(81 r_g^2) \frac{r^6}{r_0^{12}} + (72 r_g^2) \frac{1}{r_0^6} \nonumber\\
& + (24 r_g^2) \frac{1}{r_0^3} \frac{1}{r^3} + (12 r_g^2) \frac{1}{r^6} \Bigg) \frac{1}{\exp\left(\frac{r^3}{r_0^3}\right)^2}.
 \label{304}
\end{align}
These invariants are regular in the limit of $r$ tending to 0  and everywhere else. In this case, the parameter $\Bar{w}(r)$ is
related to the Kiselev varying equation of state parameter $w(r)$ as
\begin{align}
    w(r) = \frac{2(2C_4 - 3r^3)}{6C_4} -\frac{1}{3}.
\end{align}
Now, in the asymptotic limit, $w(r)$
goes as $\propto -r^3$ in contrast to the constant value for $w(r)$ related to the solution (\ref{32}). Note that the regular solutions considered in this work were obtained from an anisotropic fluid. Thus, it is possible to use the functions $B(r)$ and $w(r)$ to obtain explicitly the energy density and the radial and tangential pressures. Therefore, we can study the main energy conditions associated with our solution.
\section{Energy Conditions}
To investigate the characteristics of the matter associated with the obtained regular solutions, we can determine the conditions under which the energy density is positive, specifically, whether the weak energy conditions (WEC) are satisfied. Additionally, we aim to identify the requirements for the strong energy conditions (SEC) to hold, \textit{i.e.}, for gravity to act as an attractive force \cite{camenzind2007compact}. For more details on the energy conditions and a detailed analysis of criteria to physically reasonable non-singular
black holes, see \cite{maeda2022quest}.

The WEC is defined as 
\begin{equation} \rho \geq 0,  \label{40} \end{equation} 
and by the first two expressions of the strong energy conditions, which for an anisotropic fluid are represented by three inequalities, which we will refer to as SEC$_1$, SEC$_2$, and SEC$_3$, respectively: 
\begin{equation} 
\rho + p_t \geq 0  \quad \quad \rho + p_r \geq 0,\quad\quad \rho + p_r + 2 p_t \geq 0, \label{41} 
\end{equation} 
where SEC$_2$ is trivial, since $\rho = - p_r$. Considering the expressions for the energy density, pressures, Eqs. (\ref{40}) and (\ref{41}), we can derive two inequalities that must be satisfied for both the WEC and SEC to hold simultaneously. By substituting Equation (\ref{220}) in one the field equations, we obtain the general energy density Equation 
\begin{equation}
    \rho(r) = -\frac{C_1}{8 \pi \exp\left(\int \frac{w(r)}{r} \, dr\right)^2 r^2}.
    \label{42}
\end{equation}
Then the pressures can be obtained from equation of state 
\begin{align}
    p_r &= -\rho = \frac{C_1}{8 \pi \exp\left(\int \frac{w(r)}{r} \, dr\right)^2 r^2}, \label{43}\\
    p_t &= \Bar{w}(r)\rho(r) = -\frac{\Bar{w}(r)C_1}{8 \pi \exp\left(\int \frac{w(r)}{r} \, dr\right)^2 r^2}.
    \label{44}
\end{align}
These expressions correspond to a general solution from the Kiselev anisotropic fluid with varying equation of state parameter. In order to test conditions (\ref{40}) and (\ref{41}) we must find particular solutions for Eq. (\ref{220}). Regarding the solution associated with $\Bar{w}(r)$ defined in Eq. (\ref{31}), we obtain the Equations
\begin{align}
    \rho &=\frac{3 C_3 a^2}{8 \pi (a^2 + r^2)^{\frac{5}{2}}},
 \label{45} \\
    p_r  &= -\frac{3 C_3 a^2}{8 \pi (a^2 + r^2)^{\frac{5}{2}}},
 \label{46} \\
    p_t  &= -\frac{3 C_3 a^2 (4 a^2 - 6 r^2)}{32 \pi (a^2 + r^2)^{\frac{7}{2}}}
 \label{47},
\end{align}
these relations imply that conditions (\ref{40}) and (\ref{41}) are written in the form
\begin{align}
 \frac{3 C_3 a^2}{8 \pi (a^2 + r^2)^{\frac{5}{2}}} \geq 0, \:\:\: \frac{45 a^2 C_3 r^2}{48 \pi (a^2 + r^2)^{\frac{7}{2}}} \geq 0, 
 \label{48}
\end{align}

\begin{equation}
    -\frac{3 C_3 a^2 (4 a^2 - 6 r^2)}{16 \pi (a^2 + r^2)^{\frac{7}{2}}} \geq 0.
    \label{49}
\end{equation}
If $C_3 >0$, Eq. (\ref{49}) demands that $-(4a^2 -6r^2) \geq 0$ or \begin{equation}
    (4a^2 -6r^2) \leq 0.
    \label{50}
\end{equation}

\begin{figure}[H]
\centering
\includegraphics[scale=0.40]{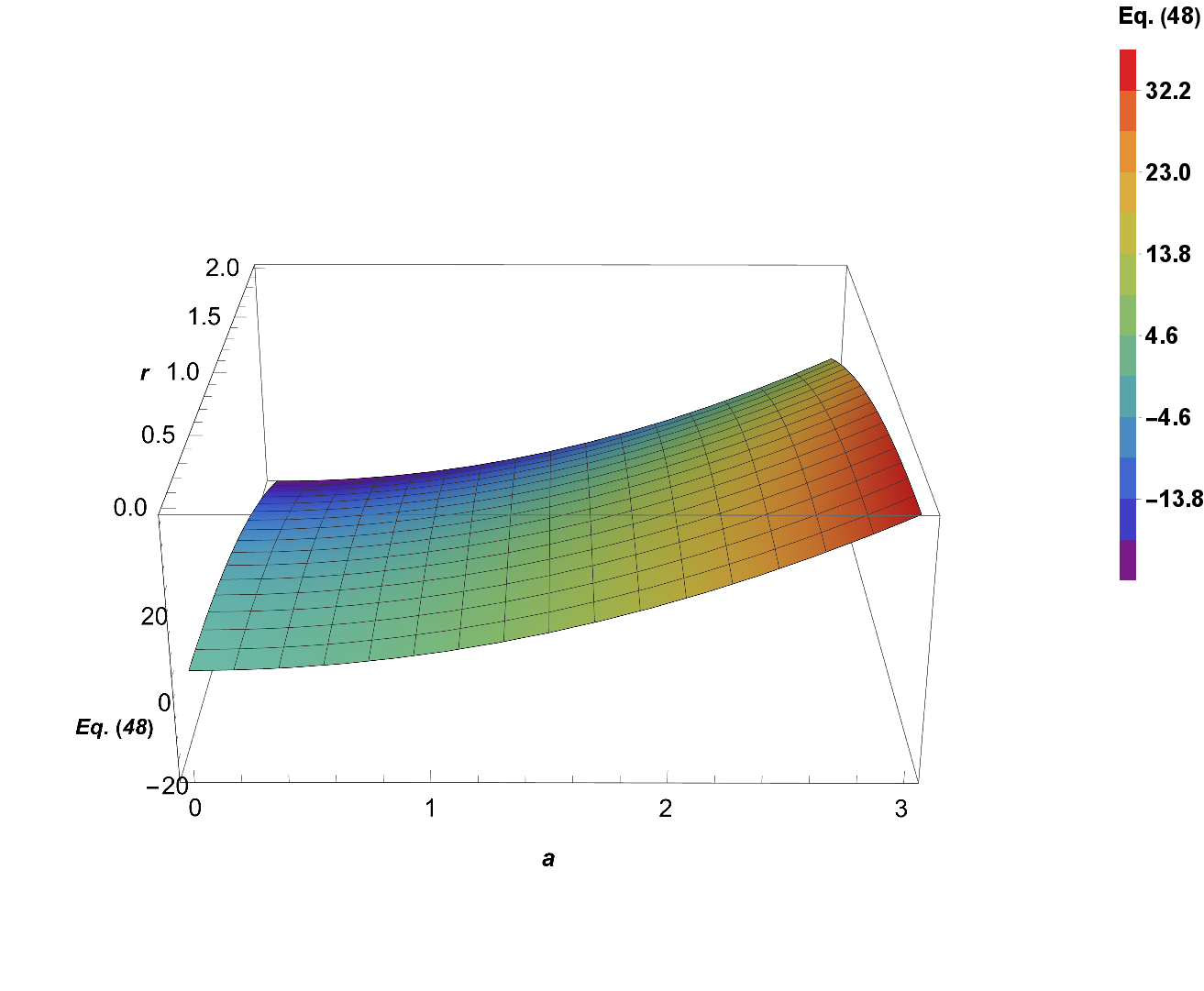} 
\caption{Condition (\ref{50}) is plotted as a function of $a$ and $r$. Negative values are associated to the regions where the SEC and WEC are satisfied.   
}
\label{f1}
\end{figure}

\begin{figure}[H]
\centering
\includegraphics[scale=0.40]{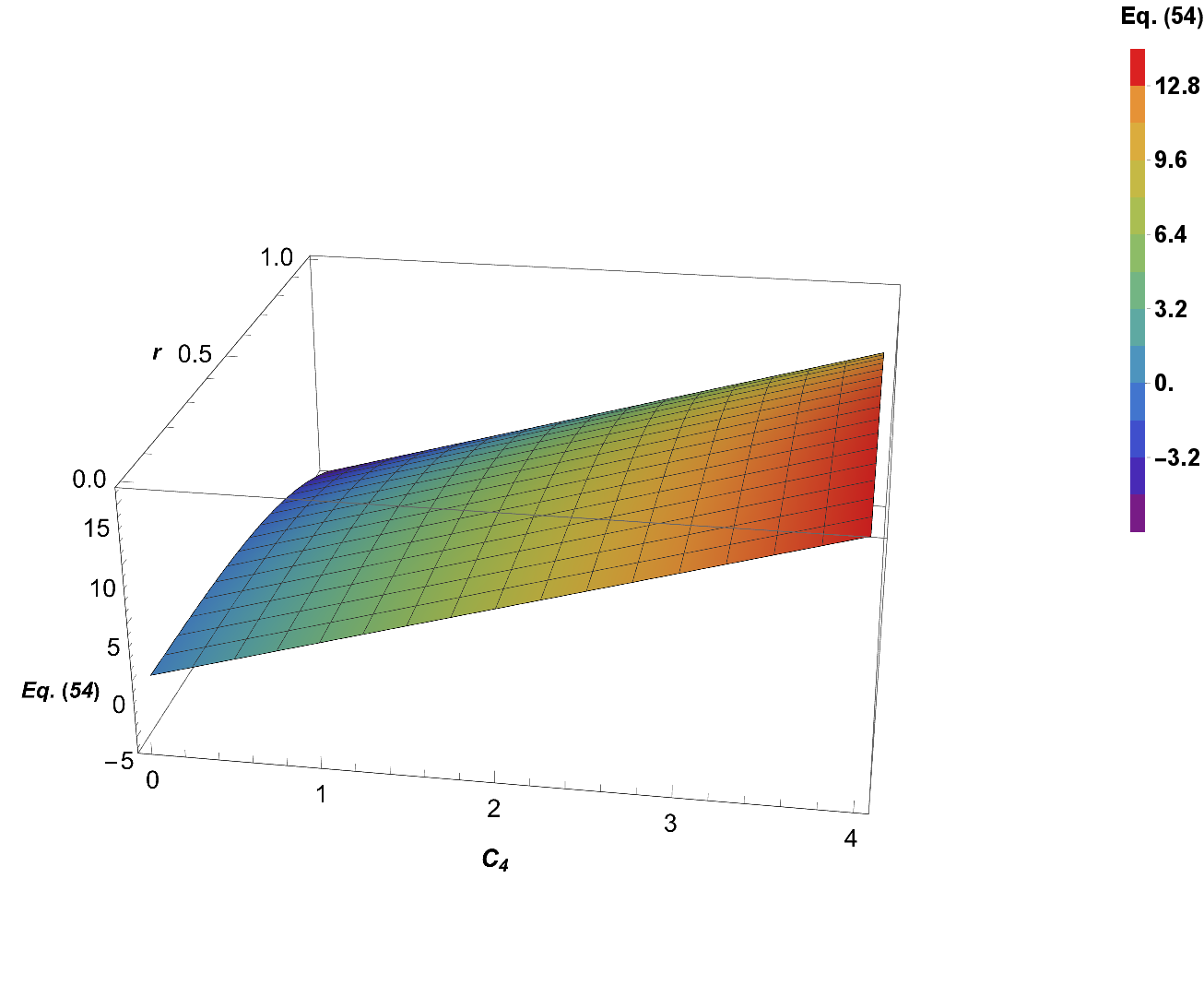} 
\caption{Condition (\ref{56}) is plotted as a function of $C_4$ and $r$. Negative values are associated to the regions where the SEC and WEC are satisfied.   
}
\label{f2}
\end{figure}

Thus, we plot in Figure \ref{f1}, the values for $r$ and $a$ in which  expression (\ref{50}) assumes negative values. 
%%%%%%%%%%%%%%%%%%%%%%%%%%%%%%%%%%%%%%%%%%
 Regarding the solution obtained using $\Bar{w}$(w) given in Eq. (\ref{38}), the expression for energy density, radial and tangential pressure are written in the form
 \begin{align}
     \rho(r) &=  -\frac{C_1\exp{\left(-\frac{r^3}{C_4}\right)}}{8\pi},
    \label{51}\\
    p_r(r) &= \frac{C_1\exp{\left(-\frac{r^3}{C_4}\right)}}{8\pi},
    \label{52}\\
    p_t(r) &=\frac{C_1 (-6 r^3 + 4 C_4) \exp\left(\frac{-r^3}{C_4}\right)}{32 C_4 \pi}.
    \label{53}
 \end{align}
Substituting these relations into conditions (\ref{40}) and (\ref{41}) we obtain
\begin{align}
   -&\frac{C_1\exp{\left(-\frac{r^3}{C_4}\right)}}{8\pi}\geq 0,\:\:\:  -\frac{6 C_1 \exp\left(\frac{-r^3}{C_4}\right) r^3}{32 C_4 \pi} \geq 0,  \label{54}\\
   & \frac{C_1 (-6 r^3 + 4 C_4) \exp\left(\frac{-r^3}{C_4}\right)}{16 C_4 \pi} \geq 0.
   \label{55}
\end{align}
If $C_1<0$, the energy conditions imply that the relation
\begin{equation}
    4C_4 - 6r^3 \leq 0,
    \label{56}
\end{equation}
must be satisfied. In Figure \ref{f2} we represent this condition.
%%%%%%%%%%%%%%%%%%%%%%%%%%%%%%%%%%%%%%%%%%
%%%%%%%%%%%%%%%%%%%%%%%%%%%%%%%%%%%%%%%%%%
%%%

As can be seen from Figures \ref{f1} and \ref{f2}, SEC and WEC can be satisfied even considering that the ``parameter'' of the equation of state depends on the radial coordinate in a non-trivial way. This suggests that the mechanism for generating new solutions from generalized equations of state has the potential to provide new solutions to the Einstein field equation with certain desired physical properties that can be associated with non-exotic matter.
\section{Concluding remarks}
In this work, we have generalized the Kiselev black hole model by introducing a radial-dependent equation of state parameter, 
$w(r)$, for the anisotropic fluid surrounding the black hole in contrast to the original Kiselev black hole solutions, which were constrained by a constant equation of state parameter. By allowing the parameter to vary with the radial coordinate, we explored novel solutions to the Einstein field equations and show the possibility of obtaining regular black hole configurations. 
Our analysis revealed that, by choosing specific forms for  $w(r)$, we were able to derive new types of geometries of black hole solutions. We showed that the curvature invariants, such as the Kretschmann scalar, remain finite at the origin, confirming the absence of singularities. This result supports the validity of the generalized model in providing regular black hole solutions offering a way to address the classical problem of singularities in general relativity.

Furthermore, we investigated the physical characteristics of the matter associated with these solutions. By studying the energy conditions, we found that the matter content surrounding the black hole remains consistent with known physical laws, even in the context of an anisotropic fluid with a varying equation of state parameter. This finding is significant because it demonstrates that the proposed model not only produces regular geometries but also preserves the physical consistency of the system.

Another important aspect of this work is the versatility of the model. The form of the function $w(r)$ can be adjusted to explore a wide range of black hole solutions, each with distinct physical properties. As a result, the model offers a flexible framework for investigating different types of black holes, including those that exhibit regular behavior. This flexibility makes the model particularly valuable for future research, as it allows the exploration new solutions depending on the form of equation of state. In addition to addressing the problem of singularities, the generalization studied may have implications for the study of black hole thermodynamics, quasinormal modes, and other dynamic phenomena. The ability to construct black hole solutions with varying equation of state parameters opens the possibility to studying the effects of different types of anisotropic matter on the evolution and stability of black holes. 

 With ongoing efforts in gravitational wave astronomy and black hole imaging, the study of regular black holes may provide new avenues for interpreting data from current and future experiments. For instance, comparing the shadows of these regular black holes with observational data from the Event Horizon Telescope or studying their gravitational wave signatures in the context of mergers could offer new tests for the existence of such exotic objects in nature. In this way, the generalization of Kiselev anisotropic fluid can offers an additional approach to study some of the issues in black hole physics. By allowing the equation of state parameter to vary with the radial coordinate, we have introduced a level of flexibility in black hole modeling, enabling the construction of physically consistent and singularity-free solutions. The results of this study can provide a foundation for future investigations into the properties of other types of geometries associated with black holes. This approach can also be explored in the context of modified gravity. In this case it may be interesting to test values for the equation of state parameter that provide solutions describing black holes in the context of extended theories of gravity. 

Thus, our findings may be are summarized as follows:
\begin{itemize}
   \item \textbf{Generalization of Kiselev Black Hole Model}: 
    Introduced a radial-dependent equation of state parameter, \( w(r) \), for the anisotropic fluid surrounding the black hole, extending the original Kiselev model which used a constant parameter. As a result the equation of state has the form $p_t(r)/\rho(r) = (3w(r) + 1)/2$ where $p_r(r) = -\rho(r) $.
    
    \item \textbf{Novel Solutions and Regular Black Holes}: 
    By allowing \( w(r) \) to vary with the radial coordinate, new solutions to the Einstein field equations were explored (See Eq. (\ref{220})), where particular solutions can leading to regular black hole configurations, free of singularities.
    
    \item \textbf{Curvature Invariants}: 
    In the case of particular solutions, the curvature invariants, like the Kretschmann scalar, were shown to remain finite at the origin, confirming the absence of singularities and validating the generalized model for regular black hole solutions.
    
    \item \textbf{Physical Consistency}: 
    Energy conditions were analyzed, demonstrating that the matter content around the black hole remained consistent, even with a varying equation of state parameter associated with Kiselev anisotropic fluid.
    
    \item \textbf{Versatility of the Model}: 
    The flexibility of the model allows for different black hole solutions based on the choice of \( w(r) \), making it useful for future research and exploration of new geometries and physical properties.

    \item \textbf{Future Research and Extended Gravity Theories}: 
    The approach can be extended to modified gravity theories, where varying equation of state parameters may describe black holes in these contexts, opening avenues for further investigation.
\end{itemize}
%%%%%%%%%%%%%%%%%%%%%%%%%%%%%%%%%%%%%%%%%%
%\begin{figure}[h]
%\centering
%\includegraphics[scale=0.27]{fig2.pdf} 
%\caption{Condition (\ref{e19b}) is plotted as a function of $w$ and $\varkappa$ considering positive values of $K$. Negative values are associated to the regions where the SEC is violated.   
%}
%\label{fig2}
%\end{figure}
%%%%%%%%%%%%%%%%%%%%%%%%%%%%%%%%%%%%%%%%%%

\acknowledgments
L.C.N.S. would like to thank FAPESC for financial support under grant 735/2024.

\bibliography{referencias_unificadas}% Produces the bibliography via BibTeX.

\begin{thebibliography}{62}%
\makeatletter
\providecommand \@ifxundefined [1]{%
 \@ifx{#1\undefined}
}%
\providecommand \@ifnum [1]{%
 \ifnum #1\expandafter \@firstoftwo
 \else \expandafter \@secondoftwo
 \fi
}%
\providecommand \@ifx [1]{%
 \ifx #1\expandafter \@firstoftwo
 \else \expandafter \@secondoftwo
 \fi
}%
\providecommand \natexlab [1]{#1}%
\providecommand \enquote  [1]{``#1''}%
\providecommand \bibnamefont  [1]{#1}%
\providecommand \bibfnamefont [1]{#1}%
\providecommand \citenamefont [1]{#1}%
\providecommand \href@noop [0]{\@secondoftwo}%
\providecommand \href [0]{\begingroup \@sanitize@url \@href}%
\providecommand \@href[1]{\@@startlink{#1}\@@href}%
\providecommand \@@href[1]{\endgroup#1\@@endlink}%
\providecommand \@sanitize@url [0]{\catcode `\\12\catcode `\$12\catcode `\&12\catcode `\#12\catcode `\^12\catcode `\_12\catcode `\%12\relax}%
\providecommand \@@startlink[1]{}%
\providecommand \@@endlink[0]{}%
\providecommand \url  [0]{\begingroup\@sanitize@url \@url }%
\providecommand \@url [1]{\endgroup\@href {#1}{\urlprefix }}%
\providecommand \urlprefix  [0]{URL }%
\providecommand \Eprint [0]{\href }%
\providecommand \doibase [0]{http://dx.doi.org/}%
\providecommand \selectlanguage [0]{\@gobble}%
\providecommand \bibinfo  [0]{\@secondoftwo}%
\providecommand \bibfield  [0]{\@secondoftwo}%
\providecommand \translation [1]{[#1]}%
\providecommand \BibitemOpen [0]{}%
\providecommand \bibitemStop [0]{}%
\providecommand \bibitemNoStop [0]{.\EOS\space}%
\providecommand \EOS [0]{\spacefactor3000\relax}%
\providecommand \BibitemShut  [1]{\csname bibitem#1\endcsname}%
\let\auto@bib@innerbib\@empty
%</preamble>
\bibitem [{\citenamefont {Hawking}(1975)}]{hawking1975particle}%
  \BibitemOpen
  \bibfield  {author} {\bibinfo {author} {\bibfnamefont {S.~W.}\ \bibnamefont {Hawking}},\ }\href@noop {} {\bibfield  {journal} {\bibinfo  {journal} {Communications in Mathematical Physics}\ }\textbf {\bibinfo {volume} {43}},\ \bibinfo {pages} {199} (\bibinfo {year} {1975})}\BibitemShut {NoStop}%
\bibitem [{\citenamefont {Penrose}(1965)}]{penrose1965gravitational}%
  \BibitemOpen
  \bibfield  {author} {\bibinfo {author} {\bibfnamefont {R.}~\bibnamefont {Penrose}},\ }\href@noop {} {\bibfield  {journal} {\bibinfo  {journal} {Physical Review Letters}\ }\textbf {\bibinfo {volume} {14}},\ \bibinfo {pages} {57} (\bibinfo {year} {1965})}\BibitemShut {NoStop}%
\bibitem [{\citenamefont {Abbott}\ \emph {et~al.}(2016)\citenamefont {Abbott}, \citenamefont {Abbott}, \citenamefont {Abbott}, \citenamefont {Zhang}, \citenamefont {Zhao}, \citenamefont {Zhou}, \citenamefont {Zhou}, \citenamefont {Zhu}, \citenamefont {Zucker}, \citenamefont {Zuraw},\ and\ \citenamefont {Zweizig}}]{waves1}%
  \BibitemOpen
  \bibfield  {author} {\bibinfo {author} {\bibfnamefont {B.~P.}\ \bibnamefont {Abbott}}, \bibinfo {author} {\bibfnamefont {R.}~\bibnamefont {Abbott}}, \bibinfo {author} {\bibfnamefont {M.}~\bibnamefont {Abbott}}, \bibinfo {author} {\bibfnamefont {Y.}~\bibnamefont {Zhang}}, \bibinfo {author} {\bibfnamefont {C.}~\bibnamefont {Zhao}}, \bibinfo {author} {\bibfnamefont {M.}~\bibnamefont {Zhou}}, \bibinfo {author} {\bibfnamefont {Z.}~\bibnamefont {Zhou}}, \bibinfo {author} {\bibfnamefont {X.~J.}\ \bibnamefont {Zhu}}, \bibinfo {author} {\bibfnamefont {M.~E.}\ \bibnamefont {Zucker}}, \bibinfo {author} {\bibfnamefont {S.~E.}\ \bibnamefont {Zuraw}}, \ and\ \bibinfo {author} {\bibfnamefont {J.}~\bibnamefont {Zweizig}} (\bibinfo {collaboration} {LIGO Scientific Collaboration and Virgo Collaboration}),\ }\href {\doibase 10.1103/PhysRevLett.116.061102} {\bibfield  {journal} {\bibinfo  {journal} {Phys. Rev. Lett.}\ }\textbf {\bibinfo {volume} {116}},\ \bibinfo {pages} {061102} (\bibinfo {year} {2016})}\BibitemShut {NoStop}%
\bibitem [{\citenamefont {Abbott}\ \emph {et~al.}(2023)\citenamefont {Abbott}, \citenamefont {Abbott}, \citenamefont {Acernese}, \citenamefont {Ackley}, \citenamefont {Adams}, \citenamefont {Adhikari}, \citenamefont {Adhikari}, \citenamefont {Adya}, \citenamefont {Affeldt}, \citenamefont {Agarwal} \emph {et~al.}}]{abbott2023gwtc}%
  \BibitemOpen
  \bibfield  {author} {\bibinfo {author} {\bibfnamefont {R.}~\bibnamefont {Abbott}}, \bibinfo {author} {\bibfnamefont {T.~D.}\ \bibnamefont {Abbott}}, \bibinfo {author} {\bibfnamefont {F.}~\bibnamefont {Acernese}}, \bibinfo {author} {\bibfnamefont {K.}~\bibnamefont {Ackley}}, \bibinfo {author} {\bibfnamefont {C.}~\bibnamefont {Adams}}, \bibinfo {author} {\bibfnamefont {N.}~\bibnamefont {Adhikari}}, \bibinfo {author} {\bibfnamefont {R.~X.}\ \bibnamefont {Adhikari}}, \bibinfo {author} {\bibfnamefont {V.~B.}\ \bibnamefont {Adya}}, \bibinfo {author} {\bibfnamefont {C.}~\bibnamefont {Affeldt}}, \bibinfo {author} {\bibfnamefont {D.}~\bibnamefont {Agarwal}},  \emph {et~al.},\ }\href@noop {} {\bibfield  {journal} {\bibinfo  {journal} {Physical Review X}\ }\textbf {\bibinfo {volume} {13}},\ \bibinfo {pages} {041039} (\bibinfo {year} {2023})}\BibitemShut {NoStop}%
\bibitem [{\citenamefont {Akiyama}\ \emph {et~al.}(2019)\citenamefont {Akiyama} \emph {et~al.}}]{black1}%
  \BibitemOpen
  \bibfield  {author} {\bibinfo {author} {\bibfnamefont {K.}~\bibnamefont {Akiyama}} \emph {et~al.} (\bibinfo {collaboration} {Event Horizon Telescope}),\ }\href {\doibase 10.3847/2041-8213/ab0ec7} {\bibfield  {journal} {\bibinfo  {journal} {Astrophys. J.}\ }\textbf {\bibinfo {volume} {875}},\ \bibinfo {pages} {L1} (\bibinfo {year} {2019})}\BibitemShut {NoStop}%
%%CITATION = ASJOA,875,L1;%%
\bibitem [{\citenamefont {Bardeen}(1968)}]{bardeen1968non}%
  \BibitemOpen
  \bibfield  {author} {\bibinfo {author} {\bibfnamefont {J.~M.}\ \bibnamefont {Bardeen}},\ }\href@noop {} {\bibfield  {journal} {\bibinfo  {journal} {Proceedings of International Conference GR5}\ } (\bibinfo {year} {1968})}\BibitemShut {NoStop}%
\bibitem [{\citenamefont {Ayon-Beato}\ and\ \citenamefont {Garcia}(1998)}]{ayon1998bardeen}%
  \BibitemOpen
  \bibfield  {author} {\bibinfo {author} {\bibfnamefont {E.}~\bibnamefont {Ayon-Beato}}\ and\ \bibinfo {author} {\bibfnamefont {A.}~\bibnamefont {Garcia}},\ }\href@noop {} {\bibfield  {journal} {\bibinfo  {journal} {Physics Letters B}\ }\textbf {\bibinfo {volume} {493}},\ \bibinfo {pages} {149} (\bibinfo {year} {1998})}\BibitemShut {NoStop}%
\bibitem [{\citenamefont {Bronnikov}(2001)}]{bronnikov2001regular}%
  \BibitemOpen
  \bibfield  {author} {\bibinfo {author} {\bibfnamefont {K.~A.}\ \bibnamefont {Bronnikov}},\ }\href@noop {} {\bibfield  {journal} {\bibinfo  {journal} {Physical Review D}\ }\textbf {\bibinfo {volume} {63}},\ \bibinfo {pages} {044005} (\bibinfo {year} {2001})}\BibitemShut {NoStop}%
\bibitem [{\citenamefont {Han}\ \emph {et~al.}(2020)\citenamefont {Han}, \citenamefont {Hu},\ and\ \citenamefont {Lan}}]{han2020thermodynamics}%
  \BibitemOpen
  \bibfield  {author} {\bibinfo {author} {\bibfnamefont {Y.-W.}\ \bibnamefont {Han}}, \bibinfo {author} {\bibfnamefont {X.-Y.}\ \bibnamefont {Hu}}, \ and\ \bibinfo {author} {\bibfnamefont {M.-J.}\ \bibnamefont {Lan}},\ }\href@noop {} {\bibfield  {journal} {\bibinfo  {journal} {The European Physical Journal Plus}\ }\textbf {\bibinfo {volume} {135}},\ \bibinfo {pages} {1} (\bibinfo {year} {2020})}\BibitemShut {NoStop}%
\bibitem [{\citenamefont {Bronnikov}(2020)}]{bronnikov2020comment}%
  \BibitemOpen
  \bibfield  {author} {\bibinfo {author} {\bibfnamefont {K.~A.}\ \bibnamefont {Bronnikov}},\ }\href@noop {} {\bibfield  {journal} {\bibinfo  {journal} {Physical Review D}\ }\textbf {\bibinfo {volume} {101}},\ \bibinfo {pages} {128501} (\bibinfo {year} {2020})}\BibitemShut {NoStop}%
\bibitem [{\citenamefont {Junior}\ \emph {et~al.}(2024)\citenamefont {Junior}, \citenamefont {Lobo},\ and\ \citenamefont {Rodrigues}}]{junior2024regular}%
  \BibitemOpen
  \bibfield  {author} {\bibinfo {author} {\bibfnamefont {J.~T. S.~S.}\ \bibnamefont {Junior}}, \bibinfo {author} {\bibfnamefont {F.~S.~N.}\ \bibnamefont {Lobo}}, \ and\ \bibinfo {author} {\bibfnamefont {M.~E.}\ \bibnamefont {Rodrigues}},\ }\href@noop {} {\bibfield  {journal} {\bibinfo  {journal} {Classical and Quantum Gravity}\ }\textbf {\bibinfo {volume} {41}},\ \bibinfo {pages} {055012} (\bibinfo {year} {2024})}\BibitemShut {NoStop}%
\bibitem [{\citenamefont {Walia}(2024)}]{walia2024exploring}%
  \BibitemOpen
  \bibfield  {author} {\bibinfo {author} {\bibfnamefont {R.~K.}\ \bibnamefont {Walia}},\ }\href@noop {} {\bibfield  {journal} {\bibinfo  {journal} {Physical Review D}\ }\textbf {\bibinfo {volume} {110}},\ \bibinfo {pages} {064058} (\bibinfo {year} {2024})}\BibitemShut {NoStop}%
\bibitem [{\citenamefont {Dolan}\ \emph {et~al.}(2024)\citenamefont {Dolan}, \citenamefont {de~Paula}, \citenamefont {Leite},\ and\ \citenamefont {Crispino}}]{dolan2024superradiant}%
  \BibitemOpen
  \bibfield  {author} {\bibinfo {author} {\bibfnamefont {S.~R.}\ \bibnamefont {Dolan}}, \bibinfo {author} {\bibfnamefont {M.~A.~A.}\ \bibnamefont {de~Paula}}, \bibinfo {author} {\bibfnamefont {L.~C.~S.}\ \bibnamefont {Leite}}, \ and\ \bibinfo {author} {\bibfnamefont {L.~C.~B.}\ \bibnamefont {Crispino}},\ }\href@noop {} {\bibfield  {journal} {\bibinfo  {journal} {Physical Review D}\ }\textbf {\bibinfo {volume} {109}},\ \bibinfo {pages} {124037} (\bibinfo {year} {2024})}\BibitemShut {NoStop}%
\bibitem [{\citenamefont {Tangphati}\ \emph {et~al.}(2024)\citenamefont {Tangphati}, \citenamefont {Youk},\ and\ \citenamefont {Ponglertsakul}}]{tangphati2024magnetically}%
  \BibitemOpen
  \bibfield  {author} {\bibinfo {author} {\bibfnamefont {T.}~\bibnamefont {Tangphati}}, \bibinfo {author} {\bibfnamefont {M.}~\bibnamefont {Youk}}, \ and\ \bibinfo {author} {\bibfnamefont {S.}~\bibnamefont {Ponglertsakul}},\ }\href@noop {} {\bibfield  {journal} {\bibinfo  {journal} {Journal of High Energy Astrophysics}\ }\textbf {\bibinfo {volume} {43}},\ \bibinfo {pages} {66} (\bibinfo {year} {2024})}\BibitemShut {NoStop}%
\bibitem [{\citenamefont {Guo}\ \emph {et~al.}(2024)\citenamefont {Guo}, \citenamefont {Xie},\ and\ \citenamefont {Miao}}]{guo2024recovery}%
  \BibitemOpen
  \bibfield  {author} {\bibinfo {author} {\bibfnamefont {Y.}~\bibnamefont {Guo}}, \bibinfo {author} {\bibfnamefont {H.}~\bibnamefont {Xie}}, \ and\ \bibinfo {author} {\bibfnamefont {Y.-G.}\ \bibnamefont {Miao}},\ }\href@noop {} {\bibfield  {journal} {\bibinfo  {journal} {Nuclear Physics B}\ }\textbf {\bibinfo {volume} {1000}},\ \bibinfo {pages} {116491} (\bibinfo {year} {2024})}\BibitemShut {NoStop}%
\bibitem [{\citenamefont {Kar}\ and\ \citenamefont {Kar}(2024)}]{kar2024novel}%
  \BibitemOpen
  \bibfield  {author} {\bibinfo {author} {\bibfnamefont {A.}~\bibnamefont {Kar}}\ and\ \bibinfo {author} {\bibfnamefont {S.}~\bibnamefont {Kar}},\ }\href@noop {} {\bibfield  {journal} {\bibinfo  {journal} {General Relativity and Gravitation}\ }\textbf {\bibinfo {volume} {56}},\ \bibinfo {pages} {52} (\bibinfo {year} {2024})}\BibitemShut {NoStop}%
\bibitem [{\citenamefont {Bronnikov}(2024)}]{bronnikov2024regular}%
  \BibitemOpen
  \bibfield  {author} {\bibinfo {author} {\bibfnamefont {K.~A.}\ \bibnamefont {Bronnikov}},\ }\href@noop {} {\bibfield  {journal} {\bibinfo  {journal} {Physical Review D}\ }\textbf {\bibinfo {volume} {110}},\ \bibinfo {pages} {024021} (\bibinfo {year} {2024})}\BibitemShut {NoStop}%
\bibitem [{\citenamefont {Bambi}\ and\ \citenamefont {Modesto}(2013)}]{bambi2013rotating}%
  \BibitemOpen
  \bibfield  {author} {\bibinfo {author} {\bibfnamefont {C.}~\bibnamefont {Bambi}}\ and\ \bibinfo {author} {\bibfnamefont {L.}~\bibnamefont {Modesto}},\ }\href@noop {} {\bibfield  {journal} {\bibinfo  {journal} {Physics Letters B}\ }\textbf {\bibinfo {volume} {721}},\ \bibinfo {pages} {329} (\bibinfo {year} {2013})}\BibitemShut {NoStop}%
\bibitem [{\citenamefont {Toshmatov}\ \emph {et~al.}(2017)\citenamefont {Toshmatov}, \citenamefont {Stuchl{\'\i}k},\ and\ \citenamefont {Ahmedov}}]{toshmatov2017generic}%
  \BibitemOpen
  \bibfield  {author} {\bibinfo {author} {\bibfnamefont {B.}~\bibnamefont {Toshmatov}}, \bibinfo {author} {\bibfnamefont {Z.}~\bibnamefont {Stuchl{\'\i}k}}, \ and\ \bibinfo {author} {\bibfnamefont {B.}~\bibnamefont {Ahmedov}},\ }\href@noop {} {\bibfield  {journal} {\bibinfo  {journal} {Physical Review D}\ }\textbf {\bibinfo {volume} {95}},\ \bibinfo {pages} {084037} (\bibinfo {year} {2017})}\BibitemShut {NoStop}%
\bibitem [{\citenamefont {Kubiz{\v{n}}{\'a}k}\ \emph {et~al.}(2022)\citenamefont {Kubiz{\v{n}}{\'a}k}, \citenamefont {Tahamtan},\ and\ \citenamefont {Svitek}}]{kubizvnak2022slowly}%
  \BibitemOpen
  \bibfield  {author} {\bibinfo {author} {\bibfnamefont {D.}~\bibnamefont {Kubiz{\v{n}}{\'a}k}}, \bibinfo {author} {\bibfnamefont {T.}~\bibnamefont {Tahamtan}}, \ and\ \bibinfo {author} {\bibfnamefont {O.}~\bibnamefont {Svitek}},\ }\href@noop {} {\bibfield  {journal} {\bibinfo  {journal} {Physical Review D}\ }\textbf {\bibinfo {volume} {105}},\ \bibinfo {pages} {104064} (\bibinfo {year} {2022})}\BibitemShut {NoStop}%
\bibitem [{\citenamefont {Kumar}\ \emph {et~al.}(2020)\citenamefont {Kumar}, \citenamefont {Kumar},\ and\ \citenamefont {Ghosh}}]{kumar2020testing}%
  \BibitemOpen
  \bibfield  {author} {\bibinfo {author} {\bibfnamefont {R.}~\bibnamefont {Kumar}}, \bibinfo {author} {\bibfnamefont {A.}~\bibnamefont {Kumar}}, \ and\ \bibinfo {author} {\bibfnamefont {S.~G.}\ \bibnamefont {Ghosh}},\ }\href@noop {} {\bibfield  {journal} {\bibinfo  {journal} {The Astrophysical Journal}\ }\textbf {\bibinfo {volume} {896}},\ \bibinfo {pages} {89} (\bibinfo {year} {2020})}\BibitemShut {NoStop}%
\bibitem [{\citenamefont {Modesto}(2004)}]{modesto2004disappearance}%
  \BibitemOpen
  \bibfield  {author} {\bibinfo {author} {\bibfnamefont {L.}~\bibnamefont {Modesto}},\ }\href@noop {} {\bibfield  {journal} {\bibinfo  {journal} {Physical Review D—Particles, Fields, Gravitation, and Cosmology}\ }\textbf {\bibinfo {volume} {70}},\ \bibinfo {pages} {124009} (\bibinfo {year} {2004})}\BibitemShut {NoStop}%
\bibitem [{\citenamefont {Gambini}\ and\ \citenamefont {Pullin}(2008)}]{gambini2008black}%
  \BibitemOpen
  \bibfield  {author} {\bibinfo {author} {\bibfnamefont {R.}~\bibnamefont {Gambini}}\ and\ \bibinfo {author} {\bibfnamefont {J.}~\bibnamefont {Pullin}},\ }\href@noop {} {\bibfield  {journal} {\bibinfo  {journal} {Physical review letters}\ }\textbf {\bibinfo {volume} {101}},\ \bibinfo {pages} {161301} (\bibinfo {year} {2008})}\BibitemShut {NoStop}%
\bibitem [{\citenamefont {Ashtekar}\ \emph {et~al.}(2023)\citenamefont {Ashtekar}, \citenamefont {Olmedo},\ and\ \citenamefont {Singh}}]{ashtekar2023regular}%
  \BibitemOpen
  \bibfield  {author} {\bibinfo {author} {\bibfnamefont {A.}~\bibnamefont {Ashtekar}}, \bibinfo {author} {\bibfnamefont {J.}~\bibnamefont {Olmedo}}, \ and\ \bibinfo {author} {\bibfnamefont {P.}~\bibnamefont {Singh}},\ }in\ \href@noop {} {\emph {\bibinfo {booktitle} {Regular Black Holes: Towards a New Paradigm of Gravitational Collapse}}}\ (\bibinfo  {publisher} {Springer},\ \bibinfo {year} {2023})\ pp.\ \bibinfo {pages} {235--282}\BibitemShut {NoStop}%
\bibitem [{\citenamefont {Modesto}(2006)}]{modesto2006loop}%
  \BibitemOpen
  \bibfield  {author} {\bibinfo {author} {\bibfnamefont {L.}~\bibnamefont {Modesto}},\ }\href@noop {} {\bibfield  {journal} {\bibinfo  {journal} {Classical and Quantum Gravity}\ }\textbf {\bibinfo {volume} {23}},\ \bibinfo {pages} {5587} (\bibinfo {year} {2006})}\BibitemShut {NoStop}%
\bibitem [{\citenamefont {Momennia}(2022)}]{momennia2022quasinormal}%
  \BibitemOpen
  \bibfield  {author} {\bibinfo {author} {\bibfnamefont {M.}~\bibnamefont {Momennia}},\ }\href@noop {} {\bibfield  {journal} {\bibinfo  {journal} {Physical Review D}\ }\textbf {\bibinfo {volume} {106}},\ \bibinfo {pages} {024052} (\bibinfo {year} {2022})}\BibitemShut {NoStop}%
\bibitem [{\citenamefont {Sharif}\ and\ \citenamefont {Javed}(2010)}]{sharif2010quantum}%
  \BibitemOpen
  \bibfield  {author} {\bibinfo {author} {\bibfnamefont {M.}~\bibnamefont {Sharif}}\ and\ \bibinfo {author} {\bibfnamefont {W.}~\bibnamefont {Javed}},\ }\href@noop {} {\bibfield  {journal} {\bibinfo  {journal} {arXiv preprint arXiv:1007.4995}\ } (\bibinfo {year} {2010})}\BibitemShut {NoStop}%
\bibitem [{\citenamefont {Kiselev}(2003)}]{kiselev}%
  \BibitemOpen
  \bibfield  {author} {\bibinfo {author} {\bibfnamefont {V.~V.}\ \bibnamefont {Kiselev}},\ }\href {\doibase 10.1088/0264-9381/20/6/310} {\bibfield  {journal} {\bibinfo  {journal} {Classical and Quantum Gravity}\ }\textbf {\bibinfo {volume} {20}},\ \bibinfo {pages} {1187} (\bibinfo {year} {2003})}\BibitemShut {NoStop}%
\bibitem [{\citenamefont {Santos}\ \emph {et~al.}(2023)\citenamefont {Santos}, \citenamefont {da~Silva}, \citenamefont {Mota}, \citenamefont {Lobo},\ and\ \citenamefont {Bezerra}}]{santos2023kiselev}%
  \BibitemOpen
  \bibfield  {author} {\bibinfo {author} {\bibfnamefont {L.~C.~N.}\ \bibnamefont {Santos}}, \bibinfo {author} {\bibfnamefont {F.~M.}\ \bibnamefont {da~Silva}}, \bibinfo {author} {\bibfnamefont {C.~E.}\ \bibnamefont {Mota}}, \bibinfo {author} {\bibfnamefont {I.~P.}\ \bibnamefont {Lobo}}, \ and\ \bibinfo {author} {\bibfnamefont {V.~B.}\ \bibnamefont {Bezerra}},\ }\href@noop {} {\bibfield  {journal} {\bibinfo  {journal} {General Relativity and Gravitation}\ }\textbf {\bibinfo {volume} {55}},\ \bibinfo {pages} {94} (\bibinfo {year} {2023})}\BibitemShut {NoStop}%
\bibitem [{\citenamefont {Konoplya}(2019)}]{shadow1}%
  \BibitemOpen
  \bibfield  {author} {\bibinfo {author} {\bibfnamefont {R.~A.}\ \bibnamefont {Konoplya}},\ }\href@noop {} {\bibfield  {journal} {\bibinfo  {journal} {Physics Letters B}\ }\textbf {\bibinfo {volume} {795}},\ \bibinfo {pages} {1} (\bibinfo {year} {2019})}\BibitemShut {NoStop}%
\bibitem [{\citenamefont {Zeng}\ and\ \citenamefont {Zhang}(2020)}]{shadow2}%
  \BibitemOpen
  \bibfield  {author} {\bibinfo {author} {\bibfnamefont {X.-X.}\ \bibnamefont {Zeng}}\ and\ \bibinfo {author} {\bibfnamefont {H.-Q.}\ \bibnamefont {Zhang}},\ }\href@noop {} {\bibfield  {journal} {\bibinfo  {journal} {The European Physical Journal C}\ }\textbf {\bibinfo {volume} {80}},\ \bibinfo {pages} {1} (\bibinfo {year} {2020})}\BibitemShut {NoStop}%
\bibitem [{\citenamefont {Abdujabbarov}\ \emph {et~al.}(2017)\citenamefont {Abdujabbarov}, \citenamefont {Toshmatov}, \citenamefont {Stuchl{\'\i}k},\ and\ \citenamefont {Ahmedov}}]{shadow3}%
  \BibitemOpen
  \bibfield  {author} {\bibinfo {author} {\bibfnamefont {A.}~\bibnamefont {Abdujabbarov}}, \bibinfo {author} {\bibfnamefont {B.}~\bibnamefont {Toshmatov}}, \bibinfo {author} {\bibfnamefont {Z.}~\bibnamefont {Stuchl{\'\i}k}}, \ and\ \bibinfo {author} {\bibfnamefont {B.}~\bibnamefont {Ahmedov}},\ }\href@noop {} {\bibfield  {journal} {\bibinfo  {journal} {International Journal of Modern Physics D}\ }\textbf {\bibinfo {volume} {26}},\ \bibinfo {pages} {1750051} (\bibinfo {year} {2017})}\BibitemShut {NoStop}%
\bibitem [{\citenamefont {He}\ \emph {et~al.}(2022)\citenamefont {He}, \citenamefont {Tao}, \citenamefont {Xue},\ and\ \citenamefont {Zhang}}]{shadow4}%
  \BibitemOpen
  \bibfield  {author} {\bibinfo {author} {\bibfnamefont {A.}~\bibnamefont {He}}, \bibinfo {author} {\bibfnamefont {J.}~\bibnamefont {Tao}}, \bibinfo {author} {\bibfnamefont {Y.}~\bibnamefont {Xue}}, \ and\ \bibinfo {author} {\bibfnamefont {L.}~\bibnamefont {Zhang}},\ }\href@noop {} {\bibfield  {journal} {\bibinfo  {journal} {Chinese Physics C}\ } (\bibinfo {year} {2022})}\BibitemShut {NoStop}%
\bibitem [{\citenamefont {Chen}\ and\ \citenamefont {Jing}(2005)}]{quasi1}%
  \BibitemOpen
  \bibfield  {author} {\bibinfo {author} {\bibfnamefont {S.}~\bibnamefont {Chen}}\ and\ \bibinfo {author} {\bibfnamefont {J.}~\bibnamefont {Jing}},\ }\href@noop {} {\bibfield  {journal} {\bibinfo  {journal} {Classical and Quantum Gravity}\ }\textbf {\bibinfo {volume} {22}},\ \bibinfo {pages} {4651} (\bibinfo {year} {2005})}\BibitemShut {NoStop}%
\bibitem [{\citenamefont {Zhang}\ and\ \citenamefont {Gui}(2006)}]{quasi2}%
  \BibitemOpen
  \bibfield  {author} {\bibinfo {author} {\bibfnamefont {Y.}~\bibnamefont {Zhang}}\ and\ \bibinfo {author} {\bibfnamefont {Y.-X.}\ \bibnamefont {Gui}},\ }\href@noop {} {\bibfield  {journal} {\bibinfo  {journal} {Classical and Quantum Gravity}\ }\textbf {\bibinfo {volume} {23}},\ \bibinfo {pages} {6141} (\bibinfo {year} {2006})}\BibitemShut {NoStop}%
\bibitem [{\citenamefont {Varghese}\ and\ \citenamefont {Kuriakose}(2009)}]{quasi3}%
  \BibitemOpen
  \bibfield  {author} {\bibinfo {author} {\bibfnamefont {N.}~\bibnamefont {Varghese}}\ and\ \bibinfo {author} {\bibfnamefont {V.~C.}\ \bibnamefont {Kuriakose}},\ }\href@noop {} {\bibfield  {journal} {\bibinfo  {journal} {General Relativity and Gravitation}\ }\textbf {\bibinfo {volume} {41}},\ \bibinfo {pages} {1249} (\bibinfo {year} {2009})}\BibitemShut {NoStop}%
\bibitem [{\citenamefont {Zhang}\ \emph {et~al.}(2007)\citenamefont {Zhang}, \citenamefont {Gui}, \citenamefont {Yu},\ and\ \citenamefont {Li}}]{quasi4}%
  \BibitemOpen
  \bibfield  {author} {\bibinfo {author} {\bibfnamefont {Y.}~\bibnamefont {Zhang}}, \bibinfo {author} {\bibfnamefont {Y.}~\bibnamefont {Gui}}, \bibinfo {author} {\bibfnamefont {F.}~\bibnamefont {Yu}}, \ and\ \bibinfo {author} {\bibfnamefont {F.}~\bibnamefont {Li}},\ }\href@noop {} {\bibfield  {journal} {\bibinfo  {journal} {General Relativity and Gravitation}\ }\textbf {\bibinfo {volume} {39}},\ \bibinfo {pages} {1003} (\bibinfo {year} {2007})}\BibitemShut {NoStop}%
\bibitem [{\citenamefont {Rayimbaev}\ \emph {et~al.}(2022)\citenamefont {Rayimbaev}, \citenamefont {Majeed}, \citenamefont {Jamil}, \citenamefont {Jusufi},\ and\ \citenamefont {Wang}}]{quasi5}%
  \BibitemOpen
  \bibfield  {author} {\bibinfo {author} {\bibfnamefont {J.}~\bibnamefont {Rayimbaev}}, \bibinfo {author} {\bibfnamefont {B.}~\bibnamefont {Majeed}}, \bibinfo {author} {\bibfnamefont {M.}~\bibnamefont {Jamil}}, \bibinfo {author} {\bibfnamefont {K.}~\bibnamefont {Jusufi}}, \ and\ \bibinfo {author} {\bibfnamefont {A.}~\bibnamefont {Wang}},\ }\href@noop {} {\bibfield  {journal} {\bibinfo  {journal} {Physics of the Dark Universe}\ }\textbf {\bibinfo {volume} {35}},\ \bibinfo {pages} {100930} (\bibinfo {year} {2022})}\BibitemShut {NoStop}%
\bibitem [{\citenamefont {Toledo}\ and\ \citenamefont {Bezerra}(2019{\natexlab{a}})}]{quasi6}%
  \BibitemOpen
  \bibfield  {author} {\bibinfo {author} {\bibfnamefont {J.~M.}\ \bibnamefont {Toledo}}\ and\ \bibinfo {author} {\bibfnamefont {V.~B.}\ \bibnamefont {Bezerra}},\ }\href@noop {} {\bibfield  {journal} {\bibinfo  {journal} {International Journal of Modern Physics D}\ }\textbf {\bibinfo {volume} {28}},\ \bibinfo {pages} {1950023} (\bibinfo {year} {2019}{\natexlab{a}})}\BibitemShut {NoStop}%
\bibitem [{\citenamefont {Saleh}\ \emph {et~al.}(2011)\citenamefont {Saleh}, \citenamefont {Bouetou},\ and\ \citenamefont {Kofane}}]{quasi7}%
  \BibitemOpen
  \bibfield  {author} {\bibinfo {author} {\bibfnamefont {M.}~\bibnamefont {Saleh}}, \bibinfo {author} {\bibfnamefont {B.~T.}\ \bibnamefont {Bouetou}}, \ and\ \bibinfo {author} {\bibfnamefont {T.~C.}\ \bibnamefont {Kofane}},\ }\href@noop {} {\bibfield  {journal} {\bibinfo  {journal} {Astrophysics and Space Science}\ }\textbf {\bibinfo {volume} {333}},\ \bibinfo {pages} {449} (\bibinfo {year} {2011})}\BibitemShut {NoStop}%
\bibitem [{\citenamefont {Tharanath}\ \emph {et~al.}(2014)\citenamefont {Tharanath}, \citenamefont {Varghese},\ and\ \citenamefont {Kuriakose}}]{quasi8}%
  \BibitemOpen
  \bibfield  {author} {\bibinfo {author} {\bibfnamefont {R.}~\bibnamefont {Tharanath}}, \bibinfo {author} {\bibfnamefont {N.}~\bibnamefont {Varghese}}, \ and\ \bibinfo {author} {\bibfnamefont {V.~C.}\ \bibnamefont {Kuriakose}},\ }\href@noop {} {\bibfield  {journal} {\bibinfo  {journal} {Modern Physics Letters A}\ }\textbf {\bibinfo {volume} {29}},\ \bibinfo {pages} {1450057} (\bibinfo {year} {2014})}\BibitemShut {NoStop}%
\bibitem [{\citenamefont {Saleh}\ \emph {et~al.}(2018{\natexlab{a}})\citenamefont {Saleh}, \citenamefont {Thomas},\ and\ \citenamefont {Kofane}}]{quasi9}%
  \BibitemOpen
  \bibfield  {author} {\bibinfo {author} {\bibfnamefont {M.}~\bibnamefont {Saleh}}, \bibinfo {author} {\bibfnamefont {B.~B.}\ \bibnamefont {Thomas}}, \ and\ \bibinfo {author} {\bibfnamefont {T.~C.}\ \bibnamefont {Kofane}},\ }\href@noop {} {\bibfield  {journal} {\bibinfo  {journal} {The European Physical Journal C}\ }\textbf {\bibinfo {volume} {78}},\ \bibinfo {pages} {1} (\bibinfo {year} {2018}{\natexlab{a}})}\BibitemShut {NoStop}%
\bibitem [{\citenamefont {Thomas}\ \emph {et~al.}(2012)\citenamefont {Thomas}, \citenamefont {Saleh},\ and\ \citenamefont {Kofane}}]{termo1}%
  \BibitemOpen
  \bibfield  {author} {\bibinfo {author} {\bibfnamefont {B.~B.}\ \bibnamefont {Thomas}}, \bibinfo {author} {\bibfnamefont {M.}~\bibnamefont {Saleh}}, \ and\ \bibinfo {author} {\bibfnamefont {T.~C.}\ \bibnamefont {Kofane}},\ }\href@noop {} {\bibfield  {journal} {\bibinfo  {journal} {General Relativity and Gravitation}\ }\textbf {\bibinfo {volume} {44}},\ \bibinfo {pages} {2181} (\bibinfo {year} {2012})}\BibitemShut {NoStop}%
\bibitem [{\citenamefont {Ghaderi}\ and\ \citenamefont {Malakolkalami}(2016)}]{termo2}%
  \BibitemOpen
  \bibfield  {author} {\bibinfo {author} {\bibfnamefont {K.}~\bibnamefont {Ghaderi}}\ and\ \bibinfo {author} {\bibfnamefont {B.}~\bibnamefont {Malakolkalami}},\ }\href@noop {} {\bibfield  {journal} {\bibinfo  {journal} {Nuclear Physics B}\ }\textbf {\bibinfo {volume} {903}},\ \bibinfo {pages} {10} (\bibinfo {year} {2016})}\BibitemShut {NoStop}%
\bibitem [{\citenamefont {Saleh}\ \emph {et~al.}(2018{\natexlab{b}})\citenamefont {Saleh}, \citenamefont {Thomas},\ and\ \citenamefont {Kofane}}]{termo3}%
  \BibitemOpen
  \bibfield  {author} {\bibinfo {author} {\bibfnamefont {M.}~\bibnamefont {Saleh}}, \bibinfo {author} {\bibfnamefont {B.~B.}\ \bibnamefont {Thomas}}, \ and\ \bibinfo {author} {\bibfnamefont {T.~C.}\ \bibnamefont {Kofane}},\ }\href@noop {} {\bibfield  {journal} {\bibinfo  {journal} {International Journal of Theoretical Physics}\ }\textbf {\bibinfo {volume} {57}},\ \bibinfo {pages} {2640} (\bibinfo {year} {2018}{\natexlab{b}})}\BibitemShut {NoStop}%
\bibitem [{\citenamefont {Toledo}\ and\ \citenamefont {Bezerra}(2019{\natexlab{b}})}]{termo4}%
  \BibitemOpen
  \bibfield  {author} {\bibinfo {author} {\bibfnamefont {J.~M.}\ \bibnamefont {Toledo}}\ and\ \bibinfo {author} {\bibfnamefont {V.~B.}\ \bibnamefont {Bezerra}},\ }\href@noop {} {\bibfield  {journal} {\bibinfo  {journal} {The European Physical Journal C}\ }\textbf {\bibinfo {volume} {79}},\ \bibinfo {pages} {1} (\bibinfo {year} {2019}{\natexlab{b}})}\BibitemShut {NoStop}%
\bibitem [{\citenamefont {Chen}\ \emph {et~al.}(2022)\citenamefont {Chen}, \citenamefont {L{\"u}tf{\"u}o{\u{g}}lu}, \citenamefont {Hassanabadi},\ and\ \citenamefont {Long}}]{termo5}%
  \BibitemOpen
  \bibfield  {author} {\bibinfo {author} {\bibfnamefont {H.}~\bibnamefont {Chen}}, \bibinfo {author} {\bibfnamefont {B.~C.}\ \bibnamefont {L{\"u}tf{\"u}o{\u{g}}lu}}, \bibinfo {author} {\bibfnamefont {H.}~\bibnamefont {Hassanabadi}}, \ and\ \bibinfo {author} {\bibfnamefont {Z.-W.}\ \bibnamefont {Long}},\ }\href@noop {} {\bibfield  {journal} {\bibinfo  {journal} {Physics Letters B}\ }\textbf {\bibinfo {volume} {827}},\ \bibinfo {pages} {136994} (\bibinfo {year} {2022})}\BibitemShut {NoStop}%
\bibitem [{\citenamefont {Tharanath}\ and\ \citenamefont {Kuriakose}(2013)}]{termo6}%
  \BibitemOpen
  \bibfield  {author} {\bibinfo {author} {\bibfnamefont {R.}~\bibnamefont {Tharanath}}\ and\ \bibinfo {author} {\bibfnamefont {V.~C.}\ \bibnamefont {Kuriakose}},\ }\href@noop {} {\bibfield  {journal} {\bibinfo  {journal} {Modern Physics Letters A}\ }\textbf {\bibinfo {volume} {28}},\ \bibinfo {pages} {1350003} (\bibinfo {year} {2013})}\BibitemShut {NoStop}%
\bibitem [{\citenamefont {Ma}\ \emph {et~al.}(2017)\citenamefont {Ma}, \citenamefont {Zhao},\ and\ \citenamefont {Ma}}]{termo7}%
  \BibitemOpen
  \bibfield  {author} {\bibinfo {author} {\bibfnamefont {M.-S.}\ \bibnamefont {Ma}}, \bibinfo {author} {\bibfnamefont {R.}~\bibnamefont {Zhao}}, \ and\ \bibinfo {author} {\bibfnamefont {Y.-Q.}\ \bibnamefont {Ma}},\ }\href@noop {} {\bibfield  {journal} {\bibinfo  {journal} {General Relativity and Gravitation}\ }\textbf {\bibinfo {volume} {49}},\ \bibinfo {pages} {1} (\bibinfo {year} {2017})}\BibitemShut {NoStop}%
\bibitem [{\citenamefont {Xu}\ \emph {et~al.}(2019)\citenamefont {Xu}, \citenamefont {Liao},\ and\ \citenamefont {Wang}}]{termo8}%
  \BibitemOpen
  \bibfield  {author} {\bibinfo {author} {\bibfnamefont {Z.}~\bibnamefont {Xu}}, \bibinfo {author} {\bibfnamefont {Y.}~\bibnamefont {Liao}}, \ and\ \bibinfo {author} {\bibfnamefont {J.}~\bibnamefont {Wang}},\ }\href@noop {} {\bibfield  {journal} {\bibinfo  {journal} {International Journal of Modern Physics A}\ }\textbf {\bibinfo {volume} {34}},\ \bibinfo {pages} {1950185} (\bibinfo {year} {2019})}\BibitemShut {NoStop}%
\bibitem [{\citenamefont {Heydarzade}\ and\ \citenamefont {Darabi}(2017)}]{heydarzade2017black}%
  \BibitemOpen
  \bibfield  {author} {\bibinfo {author} {\bibfnamefont {Y.}~\bibnamefont {Heydarzade}}\ and\ \bibinfo {author} {\bibfnamefont {F.}~\bibnamefont {Darabi}},\ }\href@noop {} {\bibfield  {journal} {\bibinfo  {journal} {Physics Letters B}\ }\textbf {\bibinfo {volume} {771}},\ \bibinfo {pages} {365} (\bibinfo {year} {2017})}\BibitemShut {NoStop}%
\bibitem [{\citenamefont {Sakti}\ \emph {et~al.}(2020)\citenamefont {Sakti}, \citenamefont {Suroso},\ and\ \citenamefont {Zen}}]{sakti_kerrnewmannutkiselev_2020}%
  \BibitemOpen
  \bibfield  {author} {\bibinfo {author} {\bibfnamefont {M.~F. A.~R.}\ \bibnamefont {Sakti}}, \bibinfo {author} {\bibfnamefont {A.}~\bibnamefont {Suroso}}, \ and\ \bibinfo {author} {\bibfnamefont {F.~P.}\ \bibnamefont {Zen}},\ }\href {\doibase 10.1016/j.aop.2019.168062} {\bibfield  {journal} {\bibinfo  {journal} {Annals of Physics}\ }\textbf {\bibinfo {volume} {413}},\ \bibinfo {pages} {168062} (\bibinfo {year} {2020})}\BibitemShut {NoStop}%
\bibitem [{\citenamefont {Morais}\ \emph {et~al.}(2022)\citenamefont {Morais}, \citenamefont {Silva}, \citenamefont {Graça},\ and\ \citenamefont {Bezerra}}]{morais_thermodynamics_2022}%
  \BibitemOpen
  \bibfield  {author} {\bibinfo {author} {\bibfnamefont {P.~H.}\ \bibnamefont {Morais}}, \bibinfo {author} {\bibfnamefont {G.~V.}\ \bibnamefont {Silva}}, \bibinfo {author} {\bibfnamefont {J.~P.~M.}\ \bibnamefont {Graça}}, \ and\ \bibinfo {author} {\bibfnamefont {V.~B.}\ \bibnamefont {Bezerra}},\ }\href {\doibase 10.1007/s10714-021-02897-x} {\bibfield  {journal} {\bibinfo  {journal} {General Relativity and Gravitation}\ }\textbf {\bibinfo {volume} {54}},\ \bibinfo {pages} {16} (\bibinfo {year} {2022})}\BibitemShut {NoStop}%
\bibitem [{\citenamefont {Gogoi}\ \emph {et~al.}(2023)\citenamefont {Gogoi}, \citenamefont {Sekhmani}, \citenamefont {Kalita}, \citenamefont {Gogoi},\ and\ \citenamefont {Bora}}]{gogoi2023joule}%
  \BibitemOpen
  \bibfield  {author} {\bibinfo {author} {\bibfnamefont {D.~J.}\ \bibnamefont {Gogoi}}, \bibinfo {author} {\bibfnamefont {Y.}~\bibnamefont {Sekhmani}}, \bibinfo {author} {\bibfnamefont {D.}~\bibnamefont {Kalita}}, \bibinfo {author} {\bibfnamefont {N.~J.}\ \bibnamefont {Gogoi}}, \ and\ \bibinfo {author} {\bibfnamefont {J.}~\bibnamefont {Bora}},\ }\href@noop {} {\bibfield  {journal} {\bibinfo  {journal} {Fortschritte Der Physik}\ }\textbf {\bibinfo {volume} {71}},\ \bibinfo {pages} {2300010} (\bibinfo {year} {2023})}\BibitemShut {NoStop}%
\bibitem [{\citenamefont {Ghosh}\ \emph {et~al.}(2024)\citenamefont {Ghosh}, \citenamefont {Islam},\ and\ \citenamefont {Maharaj}}]{ghosh_rotating_2024}%
  \BibitemOpen
  \bibfield  {author} {\bibinfo {author} {\bibfnamefont {S.~G.}\ \bibnamefont {Ghosh}}, \bibinfo {author} {\bibfnamefont {S.~U.}\ \bibnamefont {Islam}}, \ and\ \bibinfo {author} {\bibfnamefont {S.~D.}\ \bibnamefont {Maharaj}},\ }\href {\doibase 10.1088/1402-4896/ad4833} {\bibfield  {journal} {\bibinfo  {journal} {Physica Scripta}\ }\textbf {\bibinfo {volume} {99}},\ \bibinfo {pages} {065032} (\bibinfo {year} {2024})}\BibitemShut {NoStop}%
\bibitem [{\citenamefont {Saadati}\ and\ \citenamefont {Shojai}(2021)}]{saadati2021thin}%
  \BibitemOpen
  \bibfield  {author} {\bibinfo {author} {\bibfnamefont {R.}~\bibnamefont {Saadati}}\ and\ \bibinfo {author} {\bibfnamefont {F.}~\bibnamefont {Shojai}},\ }\href@noop {} {\bibfield  {journal} {\bibinfo  {journal} {Classical and Quantum Gravity}\ }\textbf {\bibinfo {volume} {38}},\ \bibinfo {pages} {135025} (\bibinfo {year} {2021})}\BibitemShut {NoStop}%
\bibitem [{\citenamefont {Javed}\ and\ \citenamefont {Alshehri}(2024)}]{javed2024impact}%
  \BibitemOpen
  \bibfield  {author} {\bibinfo {author} {\bibfnamefont {F.}~\bibnamefont {Javed}}\ and\ \bibinfo {author} {\bibfnamefont {M.~H.}\ \bibnamefont {Alshehri}},\ }\href@noop {} {\bibfield  {journal} {\bibinfo  {journal} {Annals of Physics}\ }\textbf {\bibinfo {volume} {464}},\ \bibinfo {pages} {169658} (\bibinfo {year} {2024})}\BibitemShut {NoStop}%
\bibitem [{\citenamefont {Javed}\ \emph {et~al.}(2024)\citenamefont {Javed}, \citenamefont {Shaukat}, \citenamefont {Waseem}, \citenamefont {Mustafa},\ and\ \citenamefont {Almutairi}}]{javed2024klein}%
  \BibitemOpen
  \bibfield  {author} {\bibinfo {author} {\bibfnamefont {F.}~\bibnamefont {Javed}}, \bibinfo {author} {\bibfnamefont {S.}~\bibnamefont {Shaukat}}, \bibinfo {author} {\bibfnamefont {A.}~\bibnamefont {Waseem}}, \bibinfo {author} {\bibfnamefont {G.}~\bibnamefont {Mustafa}}, \ and\ \bibinfo {author} {\bibfnamefont {B.}~\bibnamefont {Almutairi}},\ }\href@noop {} {\bibfield  {journal} {\bibinfo  {journal} {Physics of the Dark Universe}\ ,\ \bibinfo {pages} {101689}} (\bibinfo {year} {2024})}\BibitemShut {NoStop}%
\bibitem [{\citenamefont {Mota}\ \emph {et~al.}(2022)\citenamefont {Mota}, \citenamefont {Santos}, \citenamefont {da~Silva}, \citenamefont {Flores}, \citenamefont {da~Silva},\ and\ \citenamefont {Menezes}}]{santos00}%
  \BibitemOpen
  \bibfield  {author} {\bibinfo {author} {\bibfnamefont {C.~E.}\ \bibnamefont {Mota}}, \bibinfo {author} {\bibfnamefont {L.~C.~N.}\ \bibnamefont {Santos}}, \bibinfo {author} {\bibfnamefont {F.~M.}\ \bibnamefont {da~Silva}}, \bibinfo {author} {\bibfnamefont {C.~V.}\ \bibnamefont {Flores}}, \bibinfo {author} {\bibfnamefont {T.~J.~N.}\ \bibnamefont {da~Silva}}, \ and\ \bibinfo {author} {\bibfnamefont {D.~P.}\ \bibnamefont {Menezes}},\ }\href {\doibase 10.1088/1361-6382/ac5a13} {\bibfield  {journal} {\bibinfo  {journal} {Class. Quant. Grav.}\ }\textbf {\bibinfo {volume} {39}},\ \bibinfo {pages} {085008} (\bibinfo {year} {2022})}\BibitemShut {NoStop}%
\bibitem [{\citenamefont {Dymnikova}(1992)}]{dymnikova1992vacuum}%
  \BibitemOpen
  \bibfield  {author} {\bibinfo {author} {\bibfnamefont {I.~G.}\ \bibnamefont {Dymnikova}},\ }\href@noop {} {\bibfield  {journal} {\bibinfo  {journal} {General Relativity and Gravitation}\ }\textbf {\bibinfo {volume} {24}},\ \bibinfo {pages} {235} (\bibinfo {year} {1992})}\BibitemShut {NoStop}%
\bibitem [{\citenamefont {Camenzind}(2007)}]{camenzind2007compact}%
  \BibitemOpen
  \bibfield  {author} {\bibinfo {author} {\bibfnamefont {M.}~\bibnamefont {Camenzind}},\ }\href@noop {} {\emph {\bibinfo {title} {Compact objects in astrophysics}}}\ (\bibinfo  {publisher} {Springer},\ \bibinfo {year} {2007})\BibitemShut {NoStop}%
\bibitem [{\citenamefont {Maeda}(2022)}]{maeda2022quest}%
  \BibitemOpen
  \bibfield  {author} {\bibinfo {author} {\bibfnamefont {H.}~\bibnamefont {Maeda}},\ }\href@noop {} {\bibfield  {journal} {\bibinfo  {journal} {Journal of High Energy Physics}\ }\textbf {\bibinfo {volume} {2022}},\ \bibinfo {pages} {1} (\bibinfo {year} {2022})}\BibitemShut {NoStop}%
\end{thebibliography}%

\end{document}